# The Fundamental Plane in CL 0024 at $z = 0.4$: Implications for the Evolution of the Mass-to-Light Ratio [*]

Pieter G. van Dokkum, [1] and Marijn Franx [1,2] [†]
[1] *Kapteyn Astronomical Institute, P.O. Box 800, NL-9700 AV Groningen, The Netherlands*
[2] *Harvard-Smithsonian Center for Astrophysics, 60 Garden Street, Cambridge, MA 02318*

**ABSTRACT**
We present results on the Fundamental Plane of early-type galaxies in the rich cluster CL 0024 + 16 at $z=0.391$. The internal velocity dispersions of the galaxies have been measured from a 19 hour integration at the MMT. The photometric parameters of the galaxies have been measured from deep HST images. The galaxies satisfy a tight Fundamental Plane relation which is similar to that at low redshift. The scatter is 15 % in log $r_e$, also very similar to that at low redshift. The data show that massive early-type galaxies existed at $z=0.4$, and extend earlier studies of the luminosities and colors of early-types in rich clusters.

The evolution of the $M/L$ ratio is derived from the Fundamental Plane, by a comparison with Coma. The $M/L$ ratio increases by 31 ± 12 % between $z = 0.391$, and $z = 0.023$. The evolution is low when compared to models for stellar populations. The expected evolution depends on the IMF, $q_0$, and the formation redshift of the galaxies. The data are in agreement with high formation redshifts. The modeling is still uncertain, however, because of various possible biases. The most serious bias may be the progenitor bias: if the progenitors of some current day early-types are spirals at $z = 0.4$, they would not be included in the sample, and the sample would be biased towards the oldest galaxies.

More data are needed to measure the evolution of the Fundamental Plane more precisely, and its scatter. There is a hint that the form of the Fundamental Plane changes with redshift, and this needs to be determined better. Deeper samples on more clusters would be valuable. Studies of the richest nearby clusters may help to test the underlying hypothesis that the Fundamental Plane is identical in all clusters.

**Key words:** galaxies: evolution, galaxies: elliptical, galaxies: mass-to-light ratios, galaxies: structure of

## 1 INTRODUCTION

There is substantial evidence that galaxy evolution is a complex process. In contrast to the earliest ideas, involving an early formation and subsequently a smooth and regular dimming of the stellar light (e.g., Searle, Sargent, & Bagnuolo 1973), it is now believed that galaxies may evolve in an irregular way. Interactions, star bursts, and infall of material may drive the evolution of a typical galaxy. Observational evidence comes from the evolution of cluster galaxies (e.g., Butcher & Oemler 1978, 1984, Dressler & Gunn 1983, Couch & Sharples 1987), and studies of field galaxies (e.g., Broadhurst, Ellis, & Shanks 1988, Lilly et al. 1995). Similarly, hierarchical formation scenarios predict that galaxy evolution may be an irregular process (e.g., White & Frenk 1991).

Such complex evolution makes the interpretation of the observations more ambiguous, as the luminosities of galaxies can vary rapidly due to star bursts, and the morphologies of galaxies can change with time. As a result, the progenitors of current galaxies may have had different morphologies and masses.

Detailed studies of the evolution of galaxy morphology, luminosity, and mass are necessary to determine the relevance of these processes. Such work has now become possible with the high resolution imaging capability of the Hubble Space Telescope (e.g., Dressler et al. 1994), and deep spec-troscopy from the ground. An essential aspect of such studies is the measurement of the evolution of the masses and mass-to-light ratios ($M/L$) of galaxies. These measurements constrain the mass evolution of galaxies, and thereby infall and merging. They are necessary for a correct interpretation of the evolution of the luminosity function.

As the total masses of galaxies are very difficult to measure, relations such as the Tully-Fisher relation (Tully & Fisher 1977), the Faber-Jackson relation (Faber & Jackson 1976), or the Fundamental Plane (Dressler et al. 1987; Djorgovski & Davis 1987) can be used to constrain the mass evolution of galaxies. The small intrinsic scatter of these relations makes them very useful for studies of evolutionary effects.

Here we present results on the Fundamental Plane relation at high redshift. The Fundamental Plane is a relation between effective radius $r_e$, effective surface brightness $I_e$, and central velocity dispersion $\sigma$ of the form $r_e \propto \sigma^{1.24} I_e^{-0.82}$ (Bender, Burstein, & Faber 1992, Jørgensen, Franx & Kjærgaard 1995c [JFK95c]). As shown originally by Faber et al. (1987), the implication of the relation is that the $M/L$ ratio of galaxies is well-behaved, and of the form $M/L \propto r_e^{0.22} \sigma^{0.49} \propto M^{0.24}$. Recent studies indicate that the scatter in the $M/L$ for cluster galaxies is small (23%, e.g., Lucey et al. 1991, JFK95c). The small scatter makes the Fundamental Plane especially useful to measure the evolution of galaxies with redshift, as small samples can provide useful answers. Furthermore, biases due to sample selection effects are small, if the scatter is small. The study of the Fundamental Plane at higher redshift has recently become possible with high throughput, high spectral resolution spectrographs (e.g., Franx 1993a,b).

In this paper, we report on measurements of velocity dispersions and structural parameters of galaxies in the rich and concentrated cluster CL 0024+1654 at $z = 0.39$. The Fundamental Plane is derived, and the implications for the evolution of the mass-to-light ratio are discussed. Sections 2 and 3 present the spectroscopy and the photometry. Those readers who are not interested in the details of the data analysis can proceed directly to sections 4 and 5, where the Fundamental Plane is presented and discussed.

## 2 SPECTROSCOPY

Objects in the CL 0024+1654 field were observed with the Multiple Mirror Telescope (MMT) in September 1992. A mask with 14 slitlets was used. Each slitlet had a width of 1.6 arcsec, and variable length. Objects were selected on the basis of their $R$ band flux (restframe $\sim B$). The slit positions were optimized to include as many of the brightest 20 galaxies as possible, and fainter galaxies were used to fill in remaining positions. The total integration time was 70 200 s, consisting of thirteen 5 400 s exposures in three nights. One object turned out to be a star, the thirteen galaxies are listed in Table 1, along with descriptions of low resolution spectra taken by Dressler, Gunn, & Schneider (1985) [DGS]. In this paper, we will use the numbering system of DGS throughout.

### 2.1 Reduction

Each slitlet was treated as a separate, long slit spectrum. For the reduction, we primarily used the standard IRAF tasks. Some tasks were written for the removal of the cosmic ray hits.

After bias subtraction a smooth gradient was present in the residual bias frames. The peak-to-peak variation in this gradient proved to be small (0.4 %), and stable in time. We measured the gradient by fitting a low-order function to the residual of an average bias frame of each night, and we subtracted this fit from the data. The remaining variations in the individual bias frames are due to noise. Internal flatfields were used to correct for the pixel-to-pixel variations in the CCD response. The position of the slits on the chip was relatively stable during the observations: total shifts were of order $\sim 2$ pixels during each night. Several averaged internal flatfields were created for each night to minimize the effects of these shifts. Since internal flatfields were taken before and after each 5 400 s exposure, small changes in the flatfields over time (e.g., the accumulation of dust grains) could be monitored well.

Cosmic ray hits were removed in several steps. First, exposures in each night were median filtered. The median frames were subtracted from the exposures, and the residual galaxy light and the residuals of skylines were modeled with appropriate functions. Cosmic ray hits were identified automatically as significant deviations from the function fits. The results from the program were inspected by eye, and corrected if necessary. Left over cosmic rays ($\sim 10$ per exposure) and bad pixels were removed interactively. Finally, the separate exposures of each night were aligned and combined, with optimal weighting.

For each night, a wavelength solution was obtained from exposures of He-Ne-Ar lamps. The frames were transformed to a common log $\lambda$ scale. The He-Ne-Ar exposures were reduced and combined in the same way as the galaxy exposures. The response function along each slitlet was determined from twilight sky exposures. After removal of the response function systematic variations in the spatial direction were $< 1$ %. The sky spectrum was derived from the spectra at the edges of each slitlet. After sky-subtraction, the average frames of the three nights were combined to form averaged, wavelength calibrated spectra of the 14 objects.

### 2.2 Instrumental Resolution

The usual procedure to determine velocity dispersions is to compare a galaxy spectrum with a star spectrum taken through the same spectrograph with the same setup. This ensures that the instrumental resolution of the star spectrum is the same as the instrumental resolution of the galaxy spectrum. Spectroscopy of high redshift galaxies does not allow this method, as the star and galaxy spectra cannot be obtained with the same setup. For the observation of stellar spectra, the grating angle needs to be changed to cover the same rest wavelengths as the spectra of the high redshift galaxies. Since the resolution is roughly constant in Å, it will change significantly in km s$^{-1}$ (Franx 1993a,b). Furthermore, the resolution is usually dependent on the position on the CCD. Hence an absolute determination of the resolution, over the whole CCD, is necessary.

**Table 1.** Spectroscopic Parameters

| DGS[a] | BO[b] | PK[c] | Spectral Description[d] | $\sigma$ | $\delta\sigma$ | Remarks |
|---|---|---|---|---|---|---|
| 84  | 15  | 122 | Weak H,K,G,Bal abs | –   | –  | Insufficient S/N |
| 111 | 26  | 105 | E-type             | 156 | 28 |                  |
| 130 |     | 68  |                    | 243 | 22 |                  |
| 137 | 42  | 80  |                    | –   | –  | Insufficient S/N |
| 158 | 62  | 87  | E-type             | 317 | 23 |                  |
| 161 | 65  | 86  | E-type             | 275 | 20 |                  |
| 162 | 72  | 108 | E-type             | 167 | 19 |                  |
| 169 | 71  | 88  | E-type             | 342 | 25 |                  |
| 186 | 87  | 61  | E-type:            | 382 | 34 |                  |
| 194 | 96  | 85  | O II, Bal abs      | –   | –  | Insufficient S/N |
| 202 | 100 | 70  | Weak E-type        | 248 | 20 |                  |
| 218 |     |     |                    | 145 | 33 |                  |
| 229 | 125 | 64  |                    | –   | –  | Insufficient S/N |

Notes:
[a] Dressler, Gunn & Schneider (1985) [DGS] number.
[b] Butcher & Oemler (1978) number.
[c] Pickles & van der Kruit (1991) number.
[d] Description of low resolution spectra from DGS.

We determined the instrumental resolution from the He-Ne-Ar exposures, by fitting Gaussians to the emission lines. The fits were made for each slitlet, and for each observing night separately. Then, smooth functions were fit to the measured dispersions. The procedure is illustrated in Fig. 1.

The instrumental resolution proved to be approximately constant at $\sigma_{\rm instr} = 1.4$ Å (70 km s$^{-1}$ at 6000 Å) for $\lambda \gtrsim 6100$ Å. At shorter wavelengths, the instrumental resolution is larger: it rises to $\sim 150$ km s$^{-1}$ at 5500 Å. The resolution also depends on the spatial coordinate: for the top and bottom slitlets $\sigma_{\rm instr}$ is $\sim 10$ % larger than for the middle slitlet. The resolution was stable from night to night.

Since each cluster exposure was bracketed by He-Ne-Ar exposures, the resolution could be determined through the nights. The resolution proved to be stable during each night, and consistent with the results derived from the mean He-Ne-Ar exposure.

The profile of a single line is determined by the instrumental characteristics, and can deviate from a Gaussian. This is especially likely to occur when the camera is out of focus on parts of the CCD. We verified the camera focus, and the shapes of the He-Ne-Ar lines by means of two exposures taken through two Hartman masks. These masks mask out either half of the parallel beam in the spectrograph. If the camera is out of focus, then there will be an offset between the two exposures in the wavelength direction. The offsets were small ($\sim 0.6$ Å) in the spectral region that we used. Furthermore, we modeled the total instrumental resolution by taking the sum of the two separate Hartman exposures. The He-Ne-Ar lines on each Hartman exposure were fitted separately, and the instrumental resolution was now defined as the sum of two Gaussians, with different widths, and an offset. The resulting line profiles were very similar to those derived from a single Gauss fit, and the profiles did not affect the final velocity dispersions of the galaxies. This gives us confidence that the modeling is adequate.

The instrumental resolution can also be determined from sky emission lines, present in the galaxy spectra. This method has the advantage that the actual galaxy exposures are used, but suffers from the fact that many sky lines are blends. Gaussians were fit to bright lines which appeared to be no blends, using the final, summed spectra. No systematic differences between the thus determined resolution and the results from the He-Ne-Ar exposures were found. Since all determinations of the instrumental resolution give consistent values, we are confident the instrumental resolution is correctly modeled.

High spectral resolution template stars were kindly made available by R. van der Marel. The observations were done May 22, 1992, with the William Herschell Telescope (WHT) on La Palma, with the blue arm of ISIS (van der Marel et al. 1994). The spectral types of the 16 stars range from G0 to M0. The spectral resolution was measured again from comparison lamp spectra, similar to our measurements of the resolution of the MMT data. The spectral resolution is constant as a function of wavelength, but depends on the width of the slit used in the observations: $\sigma_* = 0.35$ Å (1″ slit) and $\sigma_* = 0.62$ Å (1″.8 slit). The spectra cover the spectral range 4215 – 4615 Å.

The spectra were smoothed, rebinned, and smoothed again to have identical resolution as the galaxy exposures. Since the resolution of the galaxy spectra depends on the position on the chip, a separate template star was produced for each galaxy spectrum.

We checked our method for consistency by comparing velocity dispersions obtained from template star spectra observed through the wide and the narrow slit. The differences in the output dispersion are of order 1 %, and are not systematic. Hence our procedure correctly compensates for the differences in spectral resolution.

Eight additional template stars were made available by D. Fisher. They were observed May 16, 1993, with the Kast Spectrograph on the Lick Observatory 3 m telescope. The Lick spectra have a lower spectral resolution ($\sigma_* = 1.36$ Å), but cover a larger spectral range (4215 – 5617 Å) than the WHT stellar spectra. The stars were redshifted and rebinned to mimic the MMT galaxy data. Since the instrumental resolution is comparable to the instrumental resolution of our data, no further transformations were performed.

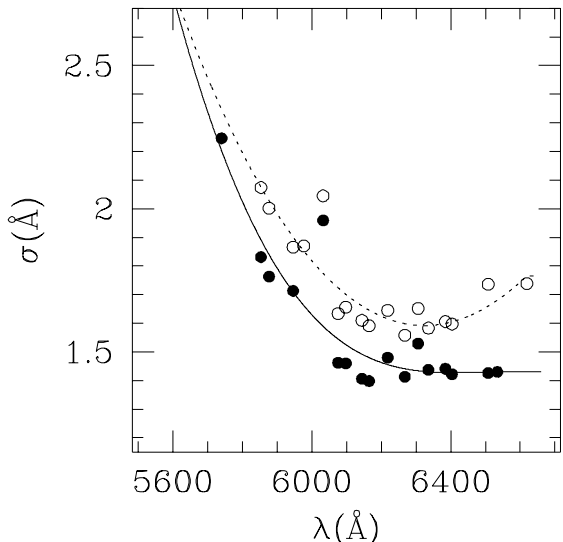

**Figure 1.** The instrumental resolution as a function of wavelength and position on the CCD. The instrumental resolution has been measured from He-Ne-Ar exposures. Solid symbols indicate the resolution for the central slitlet, open symbols the resolution for the slitlet closest to the edge of the chip. The deviant point at 6050 Å is a doublet. The resolution is highest in the centre of the chip. The lines show the smooth functions that were fit to the measurements. We used the wavelength interval from 5895 to 6400 Å to measure the velocity dispersions.

### 2.3 Fourier Fitting

Velocities and velocity dispersions were determined with the Fourier Fitting method (Franx, Illingworth, & Heckman 1989 [FIH]). This method minimizes the residuals of the fitted stellar spectrum to the galaxy spectrum

$$\chi^2 = \sum_{i=1}^{N} (G_i - [B \circ S]_i)^2 / s^2 \qquad (1)$$

where $G$ is the galaxy spectrum, $B$ is the broadening function, $S$ is the stellar spectrum, $\circ$ denotes a convolution, and $s$ is the noise in the galaxy spectrum. The fit is done in Fourier space, which speeds up the computation. The technique is described in detail in FIH.

Several other methods exist for the determination of velocities and velocity dispersions from galaxy spectra (see, e.g., Rix & White 1992). One of our objectives is to compare our results with a low redshift comparison sample. Jørgensen, Franx, & Kjærgaard (1995b) [JFK95b] have used the Fourier Fitting method for the determination of velocity dispersions of a large sample of galaxies, and they found results consistent with earlier literature data. We use the same program in a very similar setup, to limit any systematic errors.

The spectra were summed in the spatial direction to increase the S/N ratio. The three central spectra were added to give an effective aperture size of $1''\!.8 \times 1''\!.6$. The S/N decreases very rapidly at larger distances from the galaxy centers, and the effective S/N ratio cannot be improved by adding more spectra. No weighting was applied in the summation.

The Fourier program subtracted the continuum of the spectra, tapered them at the ends, and calculated the Fourier transform. The spectra were filtered with a low bandpass filter, and the fit was performed. The filter was set to give zero power at Fourier wavenumbers zero to $l_o$, and to rise slowly to unity at $2l_o$. We chose $l_o$ equivalent to 110 Å$^{-1}$. This is identical to the value used by JFK95b when expressed in units of inverse km s$^{-1}$.

Since the spectra are not flux calibrated, the shape of the continuum is partly determined by the response function of the detector. The continuum shapes of the galaxy spectra and the star spectra are somewhat different, because they were taken with different instruments. The error introduced by this difference was estimated by dividing out the slope of the stellar spectrum, and comparing the results from the fitting program. The resulting rms differences in $\sigma$ are $\sim 1\,\%$. Hence the fitting technique is insensitive to the shape of the continuum.

Various wavelength intervals for the fits were tried. The blue limit of the (redshifted) WHT and Lick star spectra is 5863 Å. Inclusion of the interval 5863 – 5895 Å resulted in a large increase in $\chi^2$ for several spectra, due to the presence of a strong sky emission line at 5890 Å. Therefore, only the part of the spectrum redward of 5895 Å (rest wavelength 4238 Å) was used in the fits. For the fits with the WHT template stars, the upper limit of the wavelength interval was determined by the red limit of the redshifted star spectrum at 6420 Å. When fitting the WHT stars, we thus used the interval 5895 – 6400 Å. The Lick stellar spectra cover a larger wavelength range. The red limit is not determined by the stellar spectrum, but by the galaxy spectrum. This limit is different for each galaxy; it is determined by the layout of the multi-aperture mask. Typically it is $\sim 6530$ Å. The results of the Fourier Fitting did not change when a red limit of 6400 Å was used.

The match between the template star spectrum and the galaxy spectrum will never be perfect, since the galaxy spectrum consists of a mix of stellar populations. In practice, template mismatch can be minimized by fitting template stars of various spectral types and using a star of the spectral type that gives the lowest $\chi^2$ residual.

Since the total number of template stars is 24, and the number of galaxies is 13, the best fitting templates could be determined reliably. The $\chi^2$ residual has a minimum at spectral types G9 – K1, and it rises gradually for cooler or hotter stars. The dependence of $\chi^2$ on spectral type is strongest for the galaxies with the highest S/N.

Three of the WHT stars and five of the Lick stars have spectral types in the range G9 – K1. The consistency of the results of the fitting program was estimated by comparing $\sigma$ and $\delta\sigma$ obtained with these template stars. The agreement between the Lick stars and the WHT stars is good. For the highest S/N spectra the distributions of the individual measurements of $\sigma$ and $\delta\sigma$ are indistinguishable; for a few galaxies the differences are up to 10 %. There is no systematic effect, i.e., the differences average out for the galaxies. The agreement gives confidence in the stability of the results. Given the fact that the instrumental resolution of the Lick stars is not precisely that of the galaxy spectra (especially blueward of 6100 Å, where the resolution of the galaxy spectra is no longer constant), the good agreement was not expected *a priori*.

Since the spectral resolution of the WHT templates is the same as the instrumental resolution of the galaxy spectra over the whole wavelength region used in the fits, we chose a good fitting WHT template (the K0 III giant HD 132737) for the final determinations of the velocities and velocity dispersions. Several galaxy spectra did not have sufficient S/N to derive a proper velocity dispersion. These galaxies were omitted from the list, as low S/N spectra produce biased estimates of the velocity dispersion (see, e.g., FIH). Fig. 2 shows the spectra of the galaxies for which reliable velocity dispersions could be determined. The position of the brightest sky emission line is indicated by bars.

Early-type galaxies have radial gradients in the velocity dispersion (e.g., Illingworth 1981, Davies et al. 1983, FIH, Jedrzejewski & Schechter 1989, van der Marel, Binney, & Davies 1990, Fisher, Illingworth, & Franx 1995). Therefore, the measured velocity dispersion through an aperture depends on the distance of the object and the aperture size. For a meaningful comparison between galaxies, the measured velocity dispersions must be corrected to a common physical aperture size. JFK95b have established aperture corrections from empirical models, based on published photometry and spectroscopy of 51 galaxies. They found that the correction to a physical aperture size of $d_{\rm corr}$ kpc can be approximated well by a power law:

$$\log \sigma_{\rm corr} = \log \sigma_{\rm obs} + 0.04 \, (\log d_{\rm obs} - \log d_{\rm corr}), \qquad (2)$$

where $d_{\rm obs}$ is the physical size in kpc of the aperture used in the observations. The correction is a simple power-law, and therefore the effective radius does not enter the expression if the aperture is corrected to a fixed physical size in kpc. We use the same normalisation as JFK95b: a $3\farcs4$ diameter circular aperture projected on a galaxy in Coma. In our observations, the effective aperture was $1\farcs6 \times 1\farcs8$. This is equivalent to a circular aperture of $1\farcs96$ diameter (JFK95b). Using Eq. 2, we find that the correction to the distance of Coma and an aperture size of $3\farcs4$ is $\sigma_{\rm corr} = 1.072 \, \sigma_{\rm obs}$, where we used a value of $q_0 = 0.5$.

Corrected velocity dispersions are listed in Table 1. The errors are the formal errors from the Fourier fitting program. Systematic errors in the velocity dispersion may be produced by two effects: spectral type mismatch, and incorrect filtering. We estimated these from the results obtained with template stars with different spectral type, and by changing the filter $l_o$ by a factor of 1.5. Each effect is on the order of 3%, and we estimated the combined systematic error at 5%.

## 3 PHOTOMETRY

Hubble Space Telescope (HST) Wide Field Camera (WFC) images of CL 0024 + 1654 were obtained from the Space Telescope Science Institute (STScI) archive. The cluster was observed through the $F702W$ filter on two occasions: October 6, 1991 and October 1, 1992. The pointings were offset by $62''$. Therefore, approximately one half of the objects were imaged in both observations. Each observation consists of six separate exposures. The images of the 1992 observation are not aligned: between the third and fourth exposure the telescope moved $3\farcs7$. Total exposure times are 11 200 seconds (1991 observation) and 13 100 seconds (1992 observation).

### 3.1 Reduction

The pipeline processing was performed at STScI. For each observation we checked whether the recommended calibration files were used. The pipeline bias frames were different from the recommended frames. We retrieved the used and recommended bias frames from the archive and subtracted them; no differences were found. Bad columns were interpolated. Cosmic rays were removed in three steps. In the first step, individual exposures of the same field were compared, and pixel values which were significantly different from the mean were rejected. Then, the exposures were averaged, and left-over hits were removed using the COSMICRAYS task in I-RAF. Finally, the images were carefully checked by eye for both left over cosmic rays and pixels which were wrongly recognized as cosmic ray hits (e.g., the central pixels of stars are sometimes marked as cosmic rays by the software).

The resulting images show a large ($\sim 20\%$) gradient in the background across the four chips. This is probably caused by non-uniformity in the throughput of the $F122M$ filter, which was used as a neutral density filter when Earth flatfields were taken (Faber 1992). Furthermore, "doughnut" like features are present ($\sim 5\%$ variations), probably images of the secondary mirror assembly (see, e.g., Ratnatunga et al. 1994).

First, we tried to remove the gradient by applying a correction flatfield, constructed from the "superskyflats" in the $F555W$ and $F785LP$ filters created by the Medium Deep Survey (MDS) team (Phillips et al. 1994; Ratnatunga et al. 1994): the MDS flatfields were divided by the stacked earth flatfield in the same filter, yielding a $F122M$ filter flatfield. These flatfields removed the "doughnuts", but did not remove the gradient satisfactorily.

Therefore, we created correction flats from the exposures themselves, by fitting a low order function to the background, carefully avoiding all objects. The correction frames were normalized to the center of WF2. These flats do not remove the "doughnuts", but do reduce the large scale fluctuations to $\sim 1.5\%$. The doughnuts were modeled and divided out separately, using the constructed $F122M$ flats.

As an independent test, we obtained images of the cluster Abell 370 from the STScI archive, taken November 11, 1992, also through the $F702W$ filter. In these observations, the same background structures are present, confirming that the source of the gradient is a flatfielding problem. After division of the Abell 370 images by our correction flats, the background was flat within 2%.

Finally, we combined the two averaged images from the 1992 observations, which were offset by $3\farcs7$. Since the shape of the HST PSF is dependent on the position on the chip, this step requires justification. We have tested the effect on galaxy parameters by refitting all (43) objects on chip 1 of the 1991 observation with PSFs offset by 18 pixels, and comparing the resulting parameters with the original ones (cf. Sect. 3.3). The average difference in $r_e$ is $0.2 \pm 0.9\,\%$ and in $I_e$ $0.3 \pm 1.7\,\%$. The difference in the product $r_e I_e^{0.8}$ is $< 0.1\,\%$. The differences in galaxy parameters do not correlate with distance from the center of the chip. Since the differences are small and are not systematic, no significant error was introduced by combining the six exposures of the 1992 observation.

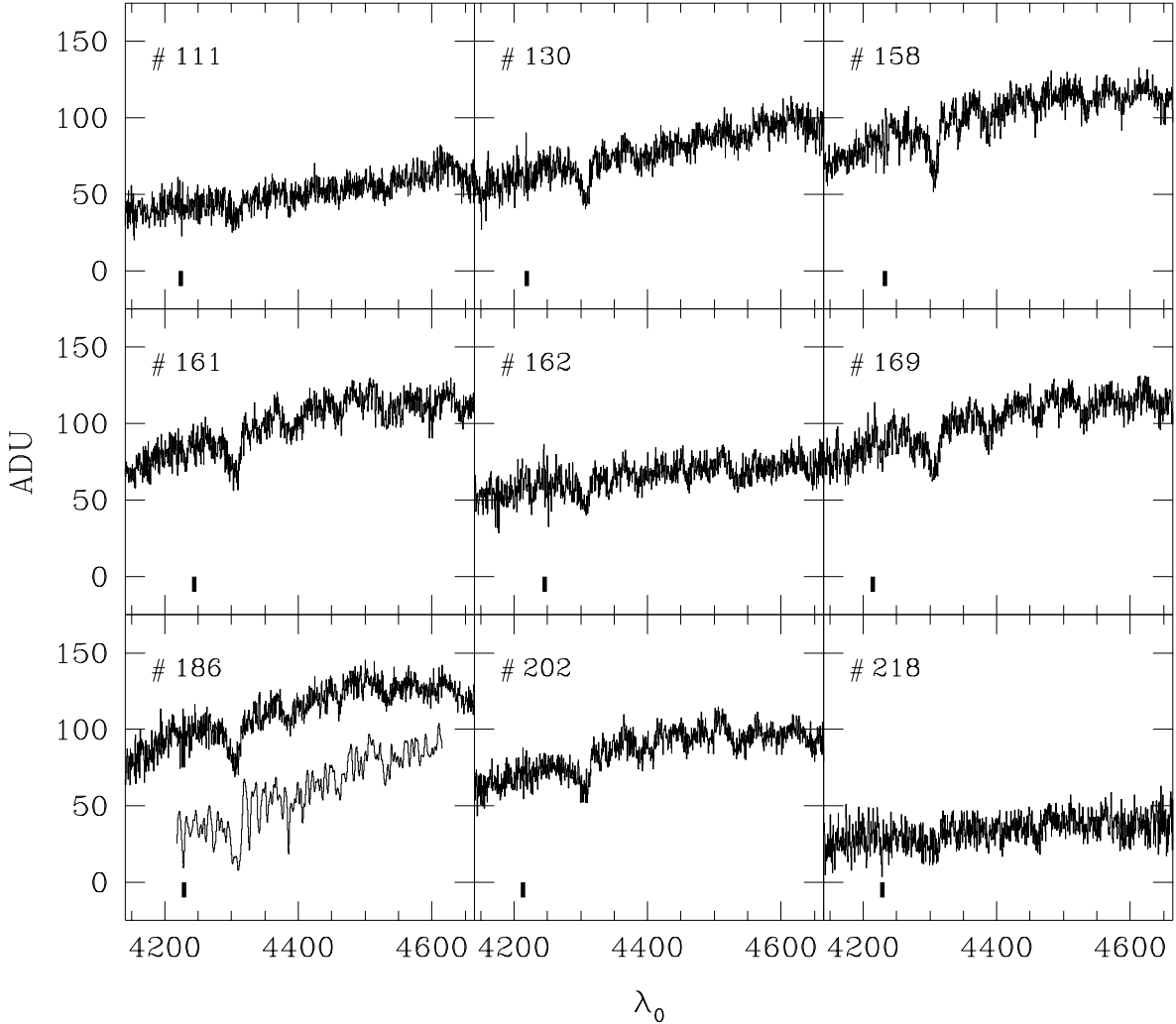

**Figure 2.** Spectra of galaxies with sufficient signal to determine internal velocity dispersions. The numbering system is from Dressler, Gunn & Schneider (1985) [DGS]. The lower left panel shows the template spectrum which was used to determine velocity dispersions. The units on the vertical axes are instrumental counts. Redshifts and velocity dispersions are listed in Table 1. Prominent spectral features are, e.g., the G-band at 4300 Å and the lines at Fe 4383, 4531, and Ca 4455 Å.

### 3.2  Zeropoints

A meaningful comparison between magnitudes of galaxies at large $z$ and magnitudes of nearby galaxies requires a transformation to a common photometric band in the restframe of the galaxies. For this purpose, the filterband needs to be shifted with the redshift of the galaxies. The $F702W$ band for the HST observations is fairly close to the redshifted $V$ band, which we denote by "$V_z$". We use a small color correction to bring the HST observations on the $V_z$ system. The steps are outlined in detail in Appendix A.

In summary, we have used groundbased data to determine a proper zeropoint for the HST instrumental magnitudes, and to determine the color correction. We find

$$V_z = m_{F702W} - (0.47 \pm 0.05)(R - I - 0.82) - (2.04 \pm 0.01), \quad (3)$$

The uncertainties in the steps are small. The systematic zeropoint uncertainty in the $V_z$ magnitude is 0.03. The noise added by the color term is small, on the order of 0.02 mag.

As a result, the method produces the true $V_z$ mag-nitudes for galaxies. It is very different from applying a $K$-correction to magnitudes taken in a fixed filter in the observers frame. As noted by many authors, such $K$-corrections produce errors due to color variations between ellipticals, and because of color evolution of galaxies with redshift. As an example, the scatter in $V - R$ for Coma ellipticals is $\sim 0.06$ magnitudes. Furthermore, the models of Worthey (1994) predict an evolution in $V - R$ of $\sim 0.10$ magnitudes between $z = 0.4$ and the present. Therefore, the various $K$-corrections in the literature can easily produce errors of order 0.1 mag. Our resulting magnitudes behave like the standard $K$-corrected magnitudes, i.e., as fluxes, and not as flux densities. This is relevant for the surface brightness dimming correction.

We determined the galactic extinction in the $R$ and $I$ band from Burstein & Heiles (1984), and Seaton (1979). We obtain an extinction of 0.09 in $R$, and 0.07 in $I$. All photometry is corrected for this extinction.

## 3.3 Analysis

### 3.3.1 Fitting Program

For an accurate measurement of length scales and central surface brightnesses it is necessary to account for the Point Spread Function (PSF). Especially for the HST, with its extended and asymmetrical PSF, an understanding of the effects of the PSF is crucial for a reliable determination of galaxy parameters.

We fitted galaxies with a code which uses the full 2D information, taking the PSF into account. The program convolves a 2D model of a galaxy with the image of the PSF, and iteratively minimizes the residual

$$\chi^2 = \sum_{i=1}^{N} w_i (I_i - [M \circ P]_i)^2, \quad (4)$$

where $I$ is the observed intensity, $M$ is the model, $P$ is the PSF, and ∘ denotes convolution. The summation is over all pixels within some fixed radius. A weighting function $w$ is applied:

$$w_i = \frac{I_i + \langle I_e \rangle}{\langle I_e \rangle}. \quad (5)$$

This weighting ensures that the weight of pixels at the effective radius is roughly half of that of the central pixels. As a stability check, we also fitted objects with a uniform weighting (i.e., $w_i = 1$) and investigated the change in derived parameters. The effect of the applied weighting function on $r_e$ and $I_e$ is $< 2\,\%$, and the effect on the FP parameter $r_e I_e^{0.8}$ $< 0.5\,\%$. This shows that the derived parameters do not depend critically on the applied weighting function.

The models are parameterized by 7 parameters: $x$ and $y$ centre position, effective radius $r_e$, surface brightness at an effective radius $I_e$, ellipticity $\epsilon$, position angle, and sky value. An additional parameter is object type: exponential, de Vaucouleurs $r^{\frac{1}{4}}$, exponential $+ r^{\frac{1}{4}}$, or a $\delta$-function, for the modeling of stars. All objects are fit simultaneously. For each object, a different PSF can be used. The advantage of fitting to the original data rather than to restored data (e.g., CLEANed data) is that the method requires a convolution instead of a deconvolution. This ensures that our method is repeatable and stable, and gives direct error estimates. Furthermore, the correlations between the parameters follow directly from the least squares fit. This is relevant here, because $r_e$ and $I_e$ are generally strongly coupled in these determinations. Fortunately, the correlation is usually parallel to the line of constant $r_e I_e^{0.8}$, and produces very little scatter in the Fundamental Plane (e.g., Jørgensen, Franx, & Kjærgaard 1993). The other advantage of this procedure is that overlapping galaxies are treated properly.

### 3.3.2 Results from the Fits

All objects in the observed CL 0024 + 1654 field with instrumental magnitudes $m_{F702W} < 24.6$ were fit. We used Tiny Tim 4.0 (Krist 1994) to calculate PSFs with a total size of $6'' \times 6''$. These were calculated for all object positions, to account for the dependence of the shape of the PSF on the position on the chip. Each galaxy was fit three times: in the first two fits, an exponential profile and an $r^{\frac{1}{4}}$ profile were fit respectively, to determine which profile type has the lowest $\chi^2$ residual. Then, a third fit was performed with the best profile type for each galaxy.

For eight galaxies in our sample an $r^{\frac{1}{4}}$ profile gives the lowest $\chi^2$ residual. Galaxy DGS 162 is best fit by an exponential profile. We tested how well we can discriminate between disk-dominated and bulge-dominated galaxies by simulations. We constructed model galaxies, with S/N and structural parameters typical for the observations. The models cover a range in inclination and bulge-to-disk ratio. We found that galaxies with bulge to disk ratios higher than 2/3 are best fit by $r^{1/4}$ law profiles. Hence, galaxies that are best fit by an $r^{\frac{1}{4}}$ profile may have significant disks, contributing up to $\sim 60\,\%$ of the light. The S/N is very high for the observations, and does not influence this result.

Fig. 3 (Plate @@) shows the images of the galaxies with measured velocity dispersions (cf. Sect. 2.3). For each galaxy, the panels show the original image, the residual from the fits to the data, and a CLEAN restoration (Högbom 1974) of the image respectively. The CLEANing was performed with a $1.5\sigma$ stopping criterion and a loop gain of 0.02. The results are stable with respect to changes in these parameters. Table 2 lists the values of $\chi^2$, and the effective parameters $r_e$ and $\mu_e$ for the galaxies. The value of $\chi^2$ is significantly higher than 1 for all galaxies. This is no surprise, as ellipticals are well known to have radial variations in ellipticity and position angle, and intensity profiles which deviate from an $r^{1/4}$ law. The galaxies with the largest $\chi^2$ are interacting galaxies (see 3.3.3).

As an independent check on our procedure, surface brightness (SB) profiles were extracted from the original and CLEANed images by integrating along ellipses with ellipticities and position angles as found by our fitting program. The profiles are shown in Fig. 4. The dotted lines indicate the result from the fit to an $r^{\frac{1}{4}}$ model; for DGS 162 the dashed line indicates the result from the fit to an exponential model.

We also verified that the direct fitting method did not miss flux within an $r_e$. To that end, we derived the fluxes of the convolved model and that of the data within an effective radius of each galaxy. The differences were small, smaller than 0.03 magnitudes.

**Figure 3.** Hubble Space Telescope Wide Field Camera images of galaxies with measured velocity dispersions. Each panel is $13'' \times 13''$. North is up, East is to the left. For each galaxy, the original data, the residual of the model fits to the data, and a CLEAN restoration is shown. Apart from DGS 186, which has a triple nucleus, the galaxies are fit well. The CLEANing tends to sharpen and emphasize features already marginally visible in the original data.

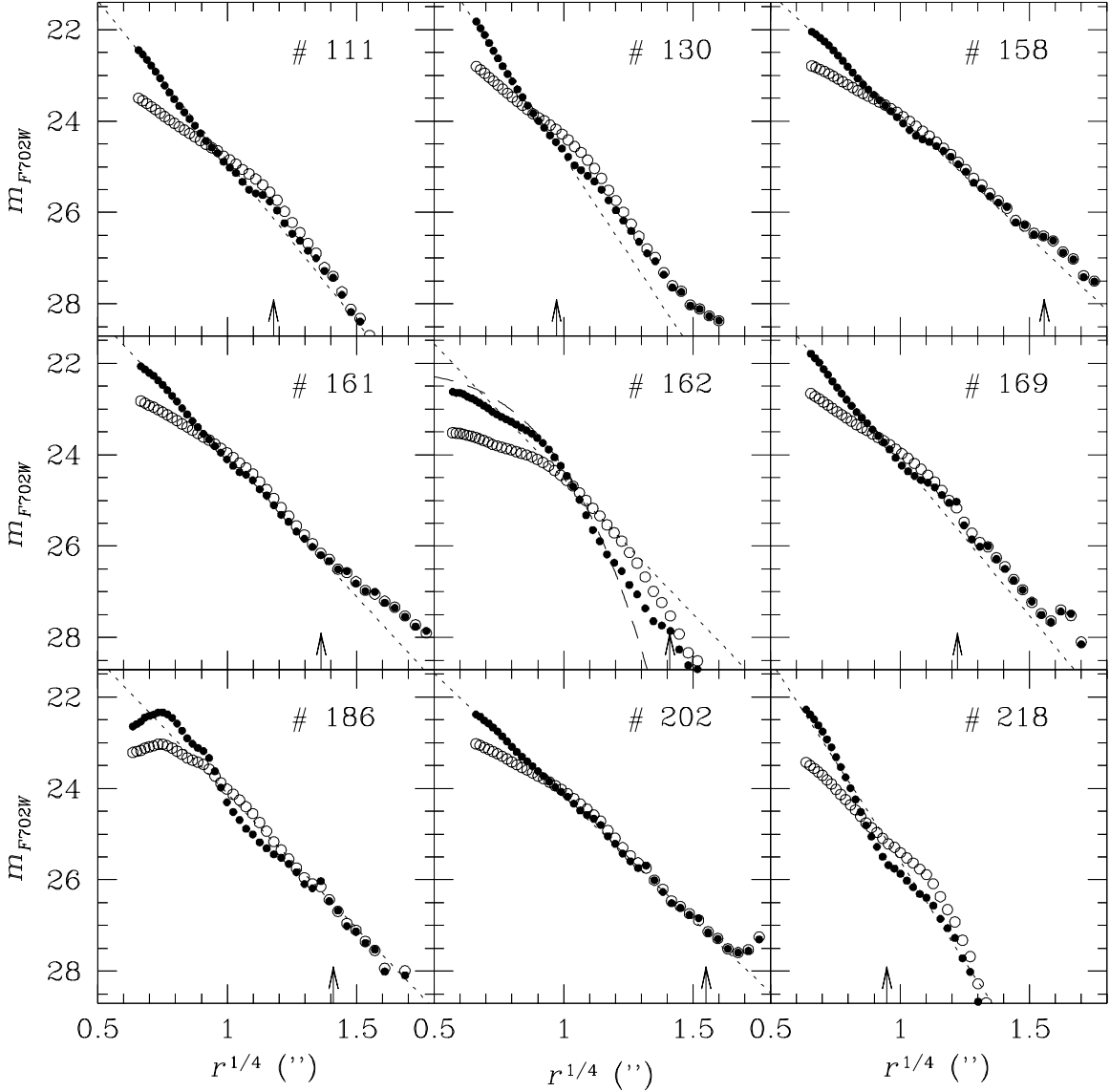

**Figure 4.** Surface brightness profiles of galaxies with measured velocity dispersions. Open circles represent the original data, filled circles a CLEAN restoration. Dotted lines indicate $r^{\frac{1}{4}}$ profiles as found by our fit to the original data. Some galaxies exhibit deviations from an $r^{\frac{1}{4}}$ profile, due to the presence of disks, or other galaxies. Galaxy DGS 162 is best fit by an exponential profile, indicated with the dashed line. Galaxy DGS 186 has a triple nucleus.

### 3.3.3 Morphologies of Galaxies with Measured Velocity Dispersions

Interestingly, several of the SB profiles depicted in Fig. 4 show significant deviations from a perfect $r^{\frac{1}{4}}$ law. The profiles are measured along ellipses with constant ellipticity, position angle and centre position. The profiles do not change when the shapes and the centers of the ellipses are allowed to vary.

We briefly discuss the morphological characteristics of the individual galaxies, on the basis of the ellipse fits, the CLEAN restorations of the images, and the residuals from the fits to the original images. We note that the distinction between ellipticals and S0's is difficult to make for lower luminosity galaxies, as face-on disks can be difficult to detect (e.g., Rix & White 1990, Jørgensen & Franx 1994).

*DGS 111* – This galaxy has the appearance of an S0 close to face-on. The inner part of the galaxy is well fit by an $r^{\frac{1}{4}}$ profile. At $r = 1''.8$ a small companion galaxy causes a bump in the profile. The profile in the outer parts ($r > 2''$) is best fit by an exponential profile, indicating this galaxy probably has a disk.

*DGS 130* – The CLEANed image, and the residual map shows a tidal tail to the North. The profile changes abruptly at $r \approx 1''.5$; the fit represents the inner part well. The outer part, containing most of the light, is fit badly. This is where the galaxy has an asymmetric extension to the North. We conclude that this galaxy is most likely interacting.

*DGS 158* – Elliptical. This is one of the four bright galaxies in the centre of the cluster. The light profile is consistent with an $r^{\frac{1}{4}}$ law.

*DGS 161* – Elliptical. This is the central galaxy of the central group in the cluster. This group consists of four galaxies, including DGS 169 and DGS 158. The profile outside of 5'' is affected by other galaxies. The inner 5'' are well fit by an $r^{\frac{1}{4}}$ law.

*DGS 162* – This is an edge-on disk galaxy. The residual image and the CLEAN image indicate dust absorption near the center and in the disk. The morphological type is S0/S. The profile is best fit by an exponential light distribution. The galaxy is red ($R - I = 0.78$); it resembles the bright, red spirals observed in nearby rich clusters. Since we intend to limit the biases as much as possible, this galaxy was treated as an elliptical (i.e., an $r^{\frac{1}{4}}$ law was fit to the profile). In Fig. 4, this fit is indicated with the dotted line.

*DGS 169* – Elliptical. The galaxy image is close to the edge of the chip, therefore, points at $r > 2''$ have large uncertainties. The inner part of the profile is well fit with an $r^{\frac{1}{4}}$ profile.

*DGS 186* – The CLEAN image of DGS 186 clearly shows the presence of three nuclei (cf. Fig 3), separated $\sim 0''.5$ (3 kpc), a characteristic of nearby Brightest Cluster Galaxies (e.g., Hoessel 1980). The galaxy is not located at the center of the cluster, but at a projected distance of 35'' (0.22 Mpc). The bright nuclei severely affect the fit to the galaxy light profile. Therefore, we do not include this galaxy in our analysis.

*DGS 202* – This galaxy appears to be an elliptical, or a close to face-on S0. It is difficult to distinguish faint ellipticals from face-on S0 galaxies. The light profile is consistent with an $r^{\frac{1}{4}}$ law.

*DGS 218* – Probably a barred S0 galaxy. The galaxy has a high ellipticity at $r \approx 0''.7$. There is a break in the light profile at $r = 0''.7$, and a position angle twist. The profile is exponential in the outer parts, consistent with the presence of a disk.

From the analysis of the HST images, it is clear that various types of galaxies are represented in our sample, ranging from a disk dominated spiral galaxy (DGS 162) to apparently 'bona fide' giant ellipticals like DGS 158. Three galaxies show evidence for disks, one galaxy is probably interacting. DGS 186 has three nuclei, and is not considered in the analysis of the Fundamental Plane.

**Table 2.** Photometric Parameters

| Iden | $^a m_r$ | Type | $\chi^2$ | $\log r_e$ '' | $\mu_e^I$ | $\mu_e^V$ | $R_c - I_c$ |
|---|---|---|---|---|---|---|---|
| 111 | 19.51 | S0 | 4.1 | 0.295 | 23.28 | 24.11 | 0.62 |
| 130 | 19.47 | Interact | 44 | $-0.048$ | 21.49 | 22.37 | 0.84 |
| 158 | 18.48 | E | 12 | 0.773 | 23.81 | 24.69 | 0.83 |
| 161 | 18.37 | E | 13 | 0.535 | 23.21 | 24.11 | 0.85 |
| 162 | 19.48 | S0/S | 12 | 0.594 | 23.94 | 24.81 | 0.78 |
| 169 | 18.57 | E | 15 | 0.348 | 22.56 | 23.45 | 0.86 |
| 186 | 19.20 | Triple | 205 | 0.601 | 23.52 | 24.40 | 0.81 |
| 202 | 19.08 | E/S0 | 14 | 0.759 | 24.08 | 24.96 | 0.82 |
| 218 | 20.55 | SB0 | 5.6 | $-0.089$ | 22.06 | 22.95 | 0.91 |

Notes:
$^a$ Gunn $r$ magnitude from Schneider, Dressler & Gunn (1986).

### 3.3.4 Internal Accuracy of the Fit Results

The fact that no systematic signal is present in the residual maps (see Fig. 3, Plate @@ ) is reassuring. Also, from Fig. 5 it is clear that the fit results generally are consistent with a CLEAN restoration. Differences between fits and data can be attributed to characteristics of the galaxies (e.g., disks or tidal features), and do not stem from systematic errors in the fitting procedure. Although the formal error in the derived parameters is small (typically $\sim 1-2$ %), we suspect the true uncertainty is larger, and primarily due to lack of knowledge of the HST PSF.

The shape of the HST PSF is position dependent due to camera vignetting, and changes over time (Faber 1992; Baggett & MacKenty 1994). Therefore, galaxies observed at different times on different chips and chip positions are ideally suited to estimate the uncertainties caused by lack of knowledge of the Point Spread Function. Approximately 40 % of the total number of objects in the cluster have been observed two times, 12 months apart, on different chips and / or at different chip positions. Therefore, we are able to derive reliable error estimates from a comparison of galaxy parameters derived from the two observations.

A total of 75 objects were observed twice. Fig. 5 shows the differences in galaxy parameters derived from the two independent sets of observations. Galaxies situated close to the edge of a WFC chip are not considered in this comparison. As expected, the differences in $r_e$ and $\mu_e$ are strongly correlated. The slope of the correlation and the spread in $r_e$ and $\mu_e$ is somewhat dependent on $r_e$; the data are grouped in three bins of $r_e$ to show this behaviour. The slope of the correlation is parallel to the line of constant $r_e I_e^{0.8}$, which enters the Fundamental Plane. There is no systematic off-set between the two exposures. With these results, we can establish the errors in the determinations of $r_e$ and $I_e$. For galaxies with $r_e > 0''.71$ the rms errors in $\log r_e$ and $\log I_e$ are 0.04 and 0.05, respectively. The rms error for $\log r_e I_e^{0.8}$, which enters the Fundamental Plane, is 0.022.

### 3.3.5 Comparison with growth curve analysis

The structural parameters of nearby early-type galaxies are usually determined through curve-of-growth fitting. We ver-ified that our technique gives similar results. To that end, we applied a curve-of-growth analysis to the CLEANed data. Except for the interacting galaxies, the resulting combination of $r_e I_e^{0.8}$ was very similar to that derived from the direct fit. The difference was generally below 5 %, except for the SB0 galaxy in the sample. The median difference in $r_e I_e^{0.8}$ was 2 %. Hence, the direct fitting method does not introduce large systematic effects.

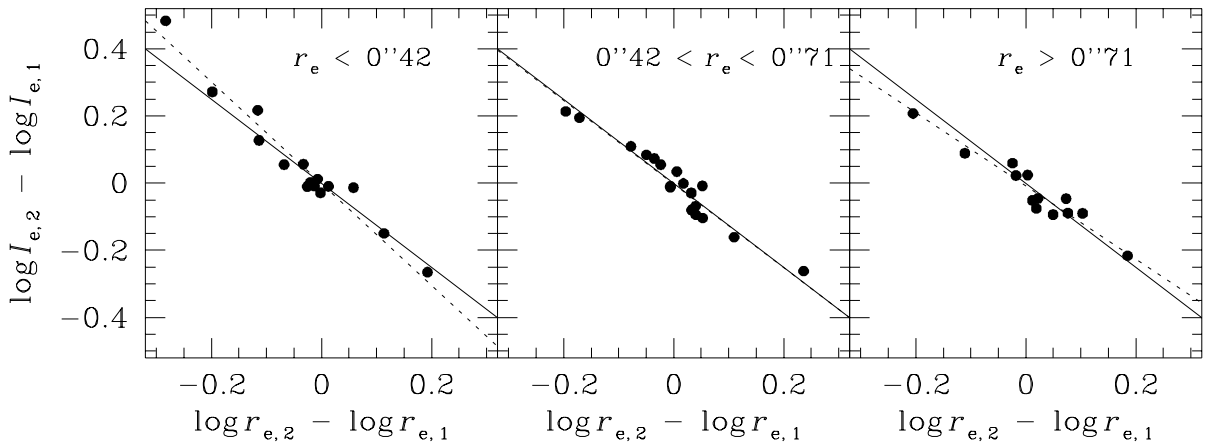

**Figure 5.** Comparison of galaxy parameters derived from the 1991 observations (labeled 1) and the 1992 observations (labeled 2). As expected, the differences in $r_e$ and $I_e$ are strongly correlated. The dotted lines are fits to the data. The uncertainty in $r_e$ and $I_e$ is lower for larger (i.e., larger $r_e$) galaxies. The observed correlations are almost parallel to the lines of constant $r_e I_e^{0.8}$ (solid lines), i.e., the Fundamental Plane parameter $r_e I_e^{0.8}$ can be determined to high accuracy. The error is estimated at 0.022 per single observation

### 3.3.6 Recovery of Flux

We also tested whether the fitting program recovered all flux, for stars and galaxies. For stars, we tested this by fitting $9'' \times 9''$ PSFs, generated with Tiny Tim 4.0, to the stars used in the calibration of the HST data (cf. Sect. A1). The aperture magnitudes were measured through an aperture with a diameter of $9''$ Fig. 6 shows the differences between the aperture magnitudes and magnitudes from the fits. The mean difference is $\langle F702W_{fit} - F702W_{9''} \rangle = 0.02 \pm 0.03$ magnitudes. From this, we infer that the results of the fitting program and the results from the stellar photometry are consistent, and the Tiny Tim construction of the PSF is a good approximation of the true PSF.

Since the wings of the HST PSF extend beyond $3''$, it may be that there are systematic errors in the derived parameters $I_e$ and/or $r_e$. We tested this by constructing model galaxies and fitting the models using PSFs with different sizes, $6''\times6''$ and $12''\times12''$ respectively. The differences in $r_e$ and $I_e$ were smaller than 3 %, whereas the combination of $r_e I_e^{0.8}$ was stable to 1 %.

In conclusion, we are confident that the parameters listed in Table 2 are not systematically different from the true galaxy parameters. The typical errors are 0.04 in $\log r_e$, 0.05 in $\log I_e$, and 0.022 in $\log r_e I_e^{0.8}$.

## 4 THE FUNDAMENTAL PLANE

In the late 1980s, it was found that early-type galaxies occupy a surface in $(r_e, I_e, \sigma)$-space, now known as the Fundamental Plane (Dressler et al. 1987; Djorgovski & Davis 1987). The scatter in the edge-on projection of the FP is small ($< 17$ % in $\log r_e$), but not entirely explained by measurement uncertainties (Lucey et al. 1991, JFK95c). The source of the intrinsic scatter in the FP is not understood at present; the residuals do not correlate with other structural parameters like ellipticity or isophotal shape (e.g., JFK95c).

The form of the Fundamental Plane is given by

$$\log r_e = \alpha \log \sigma - \beta \log I_e + \gamma. \tag{6}$$

The precise values of $\alpha$ and $\beta$ are not well known; different authors find different values, due to differences in sample selection and fitting methods. The coefficients depend on passband. However, since most data sets are not complete in more than one passband, it is often very hard to disentangle the effects of differences in passbands and sample selection.

We therefore use the coefficients measured by JFK95c, who studied 225 early-type galaxies in nearby clusters. They found $\alpha_{\text{Gunn } r} = 1.24 \pm 0.07$ and $\beta_{\text{Gunn } r} = 0.82 \pm 0.02$. We use these values throughout the analysis, and we test how the conclusions change if we use their coefficients for a bluer passband (e.g., $\alpha_{\text{Gunn } g} = 1.16 \pm 0.10$ and $\beta_{\text{Gunn } g} = 0.76 \pm 0.04$).

The Fundamental Plane implies that the $M/L$ ratio of

12  van Dokkum and Franx

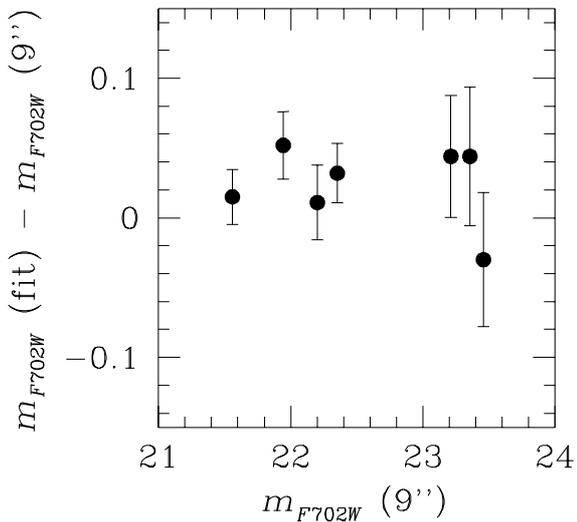

**Figure 6.** Comparison of magnitudes derived with our fitting program and magnitudes derived from aperture photometry. The fitting program used PSFs with Tiny Tim 4.0 with a size of 9'' × 9''. The aperture magnitudes were derived within 9''. The mean difference is 0.02 ± 0.03 magnitudes. We infer that the Tiny Tim PSFs are a reasonable approximation of the true PSF.

galaxies is well-behaved. Under the assumption that galaxies are a homologous family, the $M/L$ ratio scales like

$$M/L \propto \sigma^{0.49} r_e^{0.22} \propto M^{0.24} r_e^{-0.02} \qquad (7)$$

with a scatter of 23 % (Faber et al. 1987, JFK95c).

The Fundamental Plane is a good tool to study galaxy evolution, because the internal scatter is so low. It is therefore sensitive to relatively small changes. In general, evolution of the Fundamental Plane is expected, because the $M/L$ ratios of galaxies are expected to evolve with time. This is due to the evolution of the stellar population. Stellar populations with a fixed mass are expected to evolve like

$$L \propto 1/(t - t_{\rm form})^\kappa \qquad (8)$$

To first order, $\kappa = 1.3 - 0.3x$, where $x$ is the slope of the IMF (Tinsley 1980). However, $\kappa$ depends also on passband and metallicity. As an example, the models of Buzzoni (1989) and Worthey (1994) produce values of $\kappa$ between 0.6 and 0.95 in the $V$ band, for various IMFs and metallicities.

The evolution of the $M/L$ ratio implies that the zero-point of the Fundamental Plane is expected to change. Furthermore, if the luminosity evolution depends on the mass of the galaxy, the coefficients of the Fundamental Plane can change (e.g., Renzini & Ciotti 1993).

The evolution of the $M/L$ ratio is generally measured as a function of redshift. Equation (9) implies that the $M/L$ evolves in the following way

$$\ln M/L(z) = \ln M/L(0) - \kappa \left(1 + q_0 + z_{\rm form}^{-1}\right) z, \qquad (9)$$

where $z_{\rm form}$ is the formation redshift $z_{\rm form}$ of the population (Franx 1995, Franx & van Dokkum 1996a,b). This equation is valid to first order in $z$, for $q_0 \approx 0$ and high values of $z_{\rm form}$. The logarithm of the $M/L$ ratio is therefore expected to evolve in a linear way with redshift, and the slope depends on a combination of $\kappa$(IMF), $q_0$, and the formation redshift $z_{\rm form}$. In general, the $M/L$ ratio decreases with redshift, because the luminosity increases.

### 4.1 The Fundamental Plane of CL 0024

We show the Fundamental Plane for our sample of galaxies in CL 0024 in Fig. 7. The galaxies follow a tight relation, similar to nearby cluster galaxies. The scatter is 0.06 in log $r_e$, or 15 % in $r_e$. Is is comparable to the observational scatter, and similar to the scatter for Coma. Fig. 7a shows the Fundamental Plane along one of its longest axes, whereas Fig. 7b shows it along a short axis. Similar to nearby clusters, it appears tighter in one projection than in the other, but the implied scatter is the same for both projections. The small scatter in Fundamental Plane implies a scatter in age of 20-30 %, depending on the value of $\kappa$ (Renzini & Ciotti, 1993).

It is hazardous to determine the coefficients of the Fundamental Plane from the small sample presented here, because of systematic errors. Nevertheless, if we use the same algorithm as JFK95c, we find coefficients of

$$\alpha = 0.97 \pm 0.09, \beta = 0.77 \pm 0.06 \qquad (10)$$

which are not significantly different from the coefficients for Gunn $g$ derived by JFK95c. The difference in $\alpha$ is the largest, $\Delta \alpha = 0.19 \pm 0.14$. More, and deeper data are needed to measure these coefficients to higher accuracy. If future observations confirm that $\alpha$ declines with redshift, then the implication is that the $M/L$ ratio of low mass galaxies evolves faster than that of high mass galaxies. This could be due to low mass galaxies being younger (as advocated by Trager et al. 1993), to differences in the IMF (e.g., Renzini and Ciotti 1993), or unknown effects of stellar evolution.

### 4.2 The Surface Brightness Test with the Fundamental Plane

We can compare the zeropoints of the Fundamental Plane at high redshift and low redshift directly. For the low redshift sample, we use the data of Jørgensen, Franx, & Kjaergaard (1995a,b) for Coma. Their photometric parameters are based on CCD photometry, whereas the velocity dispersions come from a variety of sources. The details can be found in the Appendix. The Coma Fundamental Plane has a scatter of 16% in the $V$ band.

Fig. 8 shows the evolution of the zeropoint of the Fundamental Plane expressed in surface brightness. The zeropoint surface brightness is the surface brightness of a galaxy on the Fundamental Plane with an effective radius of 3 $h_{50}^{-1}$ kpc, and a velocity dispersion of 200 km s$^{-1}$. The decrease in surface brightness is due to the cosmological surface brightness dimming. Kjærgaard, Jørgensen, & Moles (1993) have proposed the Fundamental Plane as a method to verify the cosmological surface brightness dimming, but we note that the Cosmic Background Radiation provides a much better test. If the surface brightness would not evolve like $(1+z)^{-4}$, then the CBR would have satisfied a Planck curve for only a very short time after the photons were generated. Since it satisfies a Planck curve to 0.03 % (Mather et al. 1994), the cosmological surface brightness dimming must hold to very high accuracy.

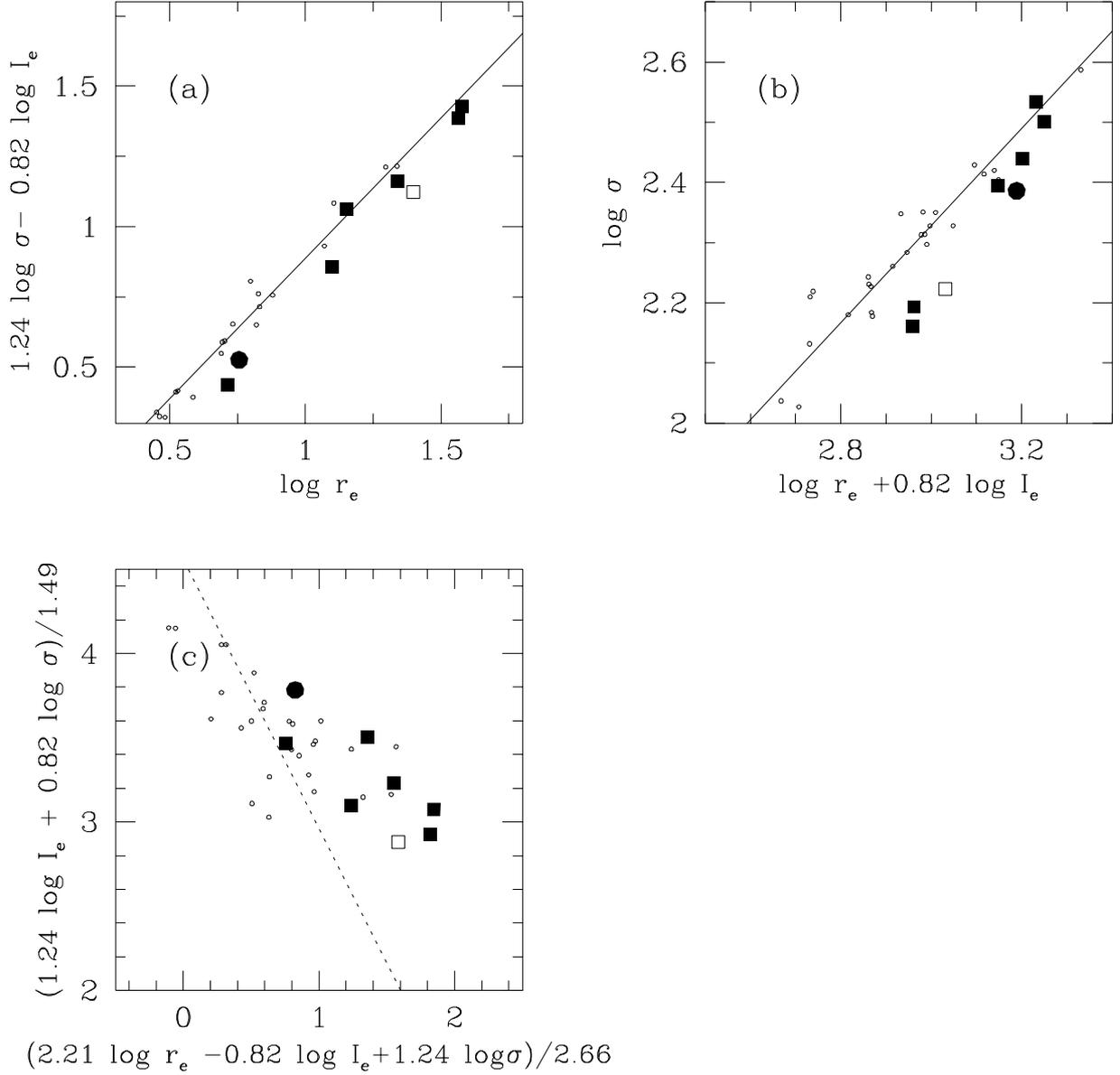

**Figure 7.** The Fundamental Plane of CL 0024+1654 in the $V$ band. a) An edge-on view of the Fundamental Plane. The two axes are two long axes of the Fundamental Plane. The large symbols are the galaxies in CL 0024, the small circles are galaxies in Coma. The open square indicates galaxy DGS 162, which is treated in the same way as the other galaxies, but probably is a spiral galaxy. The filled circle is the interacting galaxy. The solid line is the fit to the galaxies in Coma by JFK95c. The Fundamental Plane is very similar at $z = 0.4$. b) A similar edge-on Fundamental Plane. The two axes are two of the shortest axes. The Fundamental Plane appears compressed. The scatter is more prominent. c) The face-on view of the Fundamental Plane. The galaxies in CL 0024 occupy a smaller region in the Plane, which is due to the bias in favor of luminous galaxies. The dashed line is a line of constant luminosity.

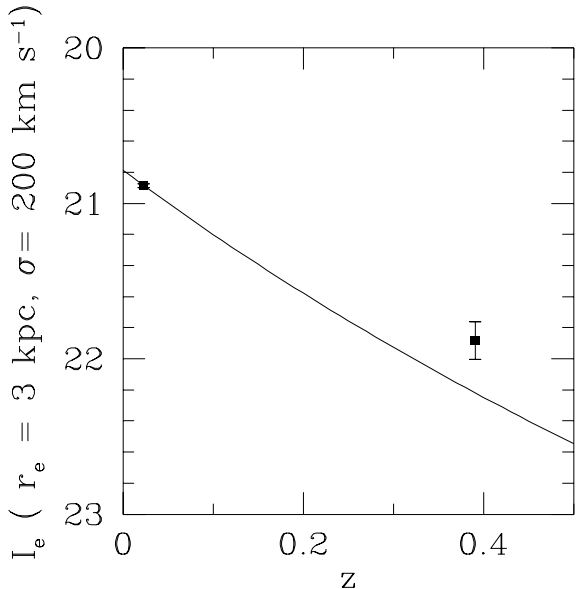

**Figure 8.** The evolution of the zeropoint of the Fundamental Plane as a function of redshift. The zeropoint is expressed as the observed $V$ surface brightness for a galaxy on the Fundamental Plane with $r_e = 3$ kpc and $\sigma = 200$ km s$^{-1}$, in units of mag arcsec$^{-2}$. The surface brightness decreases rapidly with redshift, which is due to surface brightness dimming. The drawn line is the predicted cosmological effect $I_e \propto (1+z)^{-4}$. The zeropoint for CL 0024 lies above this line, which is due to evolutionary brightening of the stellar population.

### 4.3 Evolution of the $M/L$ Ratio Indicated by the Fundamental Plane

Fig. 9 shows the $M/L$ ratios for galaxies in Coma and CL 0024 as a function of $\sigma^{0.49} r_e^{0.22}$ ($\propto M^{0.24} r_e^{-0.02}$). If galaxies evolve in a simple way, without merging or accretion, then the parameter $\sigma^{0.49} r_e^{0.22}$ remains constant, and only the $M/L$ ratio will evolve with time. The figure suggests that the tilt of the Fundamental Plane may be slightly different for CL 0024, as found earlier. The reality of the effect remains uncertain.

We can derive the mean evolution in the $M/L$ ratio by determining the offsets between both data sets. It turns out, however, that a straight fit to the data points produces an evolutionary effect which depends strongly on the coefficients of the Fundamental Plane. This is caused by the different selection of galaxies. The Coma data are dominated by much lower mass galaxies than the CL 0024 data. If the coefficients of the Fundamental Plane change, then the Coma zeropoint will be determined mostly by these low mass galaxies, whereas the CL 0024 zeropoint will be determined by the high mass galaxies. This difference in the mean mass of the galaxies results in a strong correlation between the zeropoint difference and the tilt of the lines. The difference in $\langle \log M \rangle$ for the two dataset is 0.7, equivalent to a factor of 5. The problem can be avoided by giving the low-mass galaxies in Coma low weight. All galaxies with masses lower than the lowest mass galaxy in CL 0024 were given weight 0, whereas the other galaxies were given a weight inversely proportional to the number of galaxies per mass bin. As a result, each mass bin has equal weight, both for the Coma sample, and the CL 0024 sample (which was treated in the same way). The range in masses is now very similar, and the weighted mean masses are the same.

The fits are shown as lines in Fig. 9a. The mean offset between the two datasets is 31% in the $M/L$ ratio. The offset changes very little when different coefficients are used. The coefficients for Gunn $g$ give an almost identical offset of 33%. This can be contrasted to the results when no weighting is applied: the offsets are 26% and 41% for the two sets of coefficients. The uncertainty in the offset due to random errors is very small, and it is dominated by systematic effects. The systematic effects in the structural parameters produce an uncertainty of 9 %, whereas the uncertainty due to the sample selection is estimated at 7%. The total uncertainty is therefore on the order of 12 %. Here we assumed that the systematic uncertainties can be added quadratically.

Figure 9a suggests that the slope for the galaxies in CL0024 might be somewhat different from that for Coma. This is equivalent to saying that the lower luminosity galaxies have evolved faster in their $M/L$, which is consistent with a change in the coefficient $\alpha$ in the Fundamental Plane. However, as we noted earlier, the reality of this effect is not clear, and many more data points are needed for verification.

### 4.4 Modeling the Evolution of the $M/L$ Ratio

When we apply the simple formula for a single, co-eval population to our result, we obtain a constraint of the form

$$\kappa(1 + q_0 + 1/z_{\rm form})0.37 = 0.31 \pm 0.12 \qquad (11)$$

As the models for stellar populations imply $0.6 < \kappa < 0.95$, the constraint is satisfied for a model with $z_{\rm form} = \infty$ and $q_0 = 1/2$. A model with values of $z_{\rm form} = 2$ and $q_0 = 1/2$ lies 1 $\sigma$ away, for $\kappa = 0.6$. For high values of $\kappa$, such models predict an evolution of 70%, and that is ruled out firmly.

These results, however, should be treated with caution. As pointed out by Franx (1995), and Franx and van Dokkum (1996a,b) the results from the Fundamental Plane may be biased if the fraction of early-type galaxies in clusters evolves with redshift. In such a case, the set of early-type galaxies at low redshift is not the same as the set of early-type galaxies at high redshift. Specifically, some of the early-type galaxies at $z = 0$ may be spirals at $z = 0.39$, and they should be included to derive the proper constraints on the mean formation redshift of all early-type galaxies in nearby clusters. Alternatively, we can try to measure the morphological evolution, and correct for it.

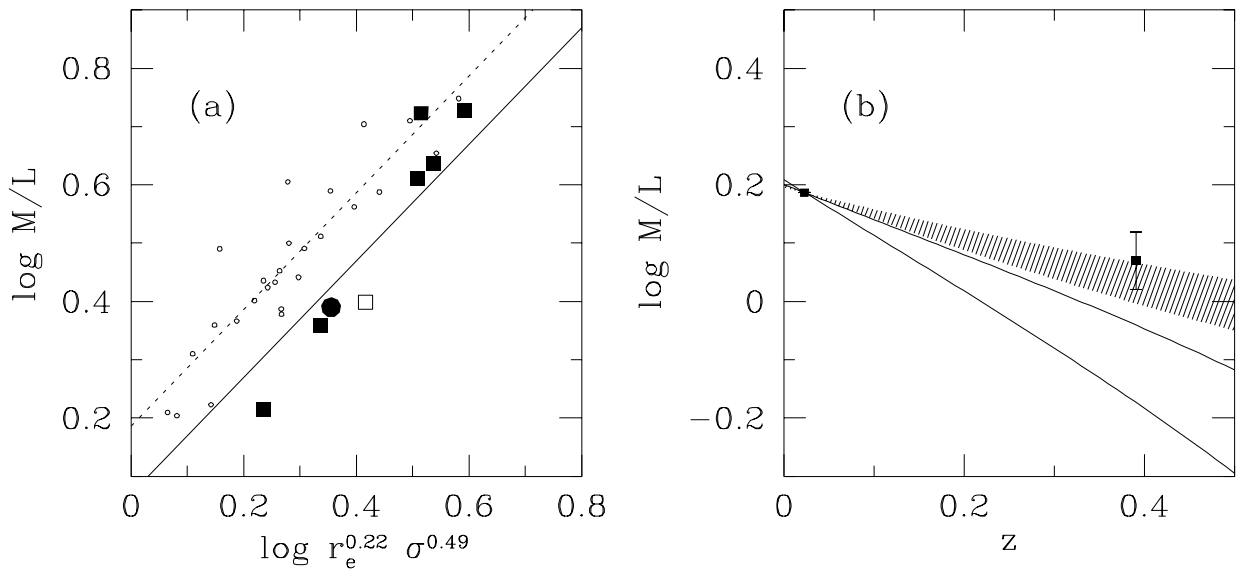

**Figure 9.** The evolution of the $M/L$ ratio of early-type galaxies in CL 0024. a) The individual measurements of the $M/L$ ratio for galaxies in CL 0024 (large symbols), and Coma (small circles). The Fundamental Plane implies that the $M/L$ ratio is proportional to the parameter plotted along the horizontal axis. The drawn line is the relation for the CL 0024 sample, the dashed line is the relation for the Coma sample. The offset between the Coma sample and the CL 0024 sample is due to the evolution of the $M/L$ ratio. The $M/L$ ratio for CL 0024 is 31 % ± 12% lower than that for Coma. The filled circle is the interacting galaxy in CL 0024, the open square is the spiral galaxy in CL 0024. There is a suggestion that the $M/L$ ratios of the low luminosity galaxies in Cl0024 are lower than expected from the Fundamental Plane. b) The evolution of the $M/L$ ratio against redshift. The $M/L$ ratio is given for a galaxy on the Fundamental Plane with $r_e = 3$ kpc and $\sigma = 200$ km s$^{-1}$, in arbitrary units. The hatched area indicates the values allowed by stellar population models with $z_{form} = \infty$ and $q_0 = 0$. The thick lines delineate the area allowed by models with $z_{form} = 1$. The data are marginally consistent with such models.

Another uncertain parameter in the equation is $\kappa$, the luminosity evolution. It depends directly on the IMF, and the IMF has not been well determined for early-type galaxies. Hence it could be that the model values of $\kappa$ are incorrect. The measurement of good optical and IR colors may help towards a resolution of this issue. The color evolution of the galaxies is also linked to the IMF, and the value of $\kappa$. For steep IMF's, $\kappa$ is small, but the color evolution will be rapid (e.g., Tinsley 1980). We notice that Stanford, Eisenhard, & Dickinson (1995) found an inconsistency between the color evolution of galaxies and the population models. The modeling of these observations needs further study.

Finally, but maybe most fundamentally, we have made the assumption that the Fundamental Plane evolves in an identical way in Coma and CL 0024. Although there is good evidence that the Fundamental Plane is very similar in Coma, and other rich nearby clusters (e.g., JFK95c), the cluster CL 0024 may not be typical of the progenitors of clusters like Coma at $z = 0.39$. Its high richness at $z = 0.39$ might imply that it will evolve into a cluster of much higher richness at $z = 0$. It is therefore valuable to probe the richest clusters at $z = 0$, to test whether the zeropoint of the Fundamental Plane is the same.

## 5 DISCUSSION AND CONCLUSIONS

We have measured photometric parameters and velocity dispersions of galaxies in the cluster CL 0024+1654 at $z$=0.39. The Fundamental Plane at this redshift is similar to the Fundamental Plane for nearby galaxies. The scatter is small, and dominated by measurement errors. Since the scatter is expected to increase with redshift, due to the fact that fractional age differences between galaxies become larger, this is rather surprising. More data are needed to measure the shape and scatter of the Fundamental Plane in greater detail. The low number of galaxies in our sample makes a good determination of these parameters difficult. Furthermore, the current sample is biased towards the most massive galaxies.

The data demonstrate that massive ellipticals existed at $z$=0.39 which are very similar to those in nearby clusters. The formation of these galaxies must have occured at significantly higher redshift. This conclusion is in agreement with earlier studies on the colors and color magnitude relation (e.g., Bower, Lucey, & Ellis 1992). The analysis shows a weak evolution of the $M/L$ ratio of the galaxies, on the order of 31% ± 12%. Such evolution is expected from the evolution of stellar populations in general. The speed of the evolution depends on the IMF, the formation redshift of the galaxy, and $q_0$. Our low evolution of the $M/L$ ratio is difficult to reconcile with a low formation redshift. However, large uncertainties in the interpretation remain, relating to the IMF, and the progenitor problem: since it is not known what the progenitors of present-day ellipticals look like at $z=0.39$, it may be that we are selecting preferentially old progenitor galaxies at $z=0.39$ and then conclude that they are old. Detailed studies of the morphological evolution of cluster galaxies is needed to address that issue.

Our results can be compared to other measurements of the luminosity evolution in rich clusters. Aragon-Salamanca et al. (1993) derived the evolution of the $K$ band luminosity function of rich clusters. They found no significant evolution. Their errors are large enough, however, to be consistent with the current observations. More such data would be valuable. The luminosity function does not necessarily evolve in the same fashion as the $M/L$ ratio. If dissipationless merging occurs frequently, then its effect on the luminosities is much larger than on the $M/L$ ratio. The combined measurement of the evolution of the luminosity function, and the $M/L$ ratio can constrain such merging.

We can also compare our results to the evolution of the luminosity function of field galaxies. Lilly et al. (1995) and Ellis et al. (1995) find little evolution of the luminosity function at high luminosities for $z=0.5$. This is consistent with our results. Furthermore, Lilly et al. analyzed their red galaxies separately, and found small evolution for the red sample as a whole. These results are in qualitative agreement with the current result. A passive evolution of 1 magnitude, or higher, at $z=0.5$ appears to be inconsistent with all results.

It may be more difficult to reconcile the low luminosity evolution with the low, or absent color evolution of early-type galaxies at $z=0.4$ (e.g., Aragon-Salamanca et al. 1993, Rakos & Schombert 1995). Stellar population models predict rapid color evolution when the luminosity evolution is low, and vice versa (Tinsley 1980).

The datasets therefore agree on a qualitative level, in the sense that evolution is small, but more data are needed to quantify all evolutionary effects more accurately. Data are needed on more galaxies in similar clusters to determine the scatter and form of the Fundamental Plane in greater detail. Such data can be used to confirm the hint that the low mass galaxies have evolved faster than high mass galaxies. Furthermore, it will be of interest to study poorer clusters at higher redshifts, to test for a dependence of the evolution on cluster properties. Finally, the comparison sample at low redshift needs to be extended towards the richest clusters.

## 6 ACKNOWLEDGMENTS

The staff of KPNO and MMT are thanked for their assistance with the observations. Dan Fabricant helped with the fabrication of the aperture masks. Roeland van der Marel and David Fisher are thanked for making their data available in digital form. This research was partly funded by Hubble Fellowship grant HF-1016.01.91A and a grant from the University of Groningen. The comments of the referee, Roger Davies, helped to improve the text.

## APPENDIX A: TRANSFORMATION FROM $M_{F702W}$ MAGNITUDES TO REDSHIFTED $V$ MAGNITUDES

### A1 Zeropoints

The HST $F702W$ zeropoint has an uncertainty of order 0.1 magnitudes (Faber 1992), primarily due to the flatfielding problems (cf. Sect. 3.1) and the uncertainty in the contamination correction. Accurate zeropoints (derived by the Medium Deep Survey team; Phillips et al. 1994) exist only for the $F555W$ and $F785LP$ filters. Hence, we used ground based observations of CL 0024 + 1654 to derive the $F702W$ zeropoint and the transformation to Cousins $R$ and $I$ magnitudes.

The observations were done on December 12, 1991, on the Kitt Peak National Observatory 2.1 m telescope, in the Cousins $R$ and $I$ bands. The integration times were 1800 s in $I$ and 300 s in $R$. The standard photometric reduction was applied, using the IRAF package. The photometric zeropoints and color terms were calculated with the PHOTCAL package. Fluxes of 19 standard stars (Landolt 1992) were measured through a 9″ diameter aperture. The color terms were small, and the published magnitudes could be reproduced to an rms accuracy of 0.02 magnitude.

The HST data of CL 0024 + 1654 were smoothed to the same resolution as the ground based data. The magnitudes of the stars and galaxies were measured through a 9″ aperture. We define $F702W$ instrumental magnitudes:

$$m_{F702W} = 25 - 2.5 \log \left( \frac{\text{ADU}}{\text{exptime}} \right). \quad (A1)$$

A transformation of the form

$$I = m_{F702W} - \alpha(R - I - \langle R - I \rangle_{\text{gal}}) + \beta \quad (A2)$$

was derived, where $\langle R - I \rangle_{\text{gal}} = 0.82$. A least squares fit to the data gave the following transformation:

$$I = m_{F702W} - (0.70 \pm 0.05)(R - I - 0.82) - (2.92 \pm 0.01). (A3)$$

The residuals from the fit are shown in Fig. A1. The scatter is consistent with the measurement errors. This transformation can be compared to the transformations of Harris et al. (1991) between the HST filter system and the Kron-Cousins filter system, and to the in-flight $F702W$ zeropoint derived in Faber (1992) from observations of $\omega$ Cen. We derived $V - I = 2.13(R - I)$ from the observed standard stars. Hence the equivalent expression to Eq. A3 is given by

$$I = F702W - 0.72(R - I - 0.82) - 2.95. \quad (A4)$$

The $R - I$ colors of the galaxies fall into a narrow range: $\langle R - I \rangle = 0.82 \pm 0.06$. Therefore, our transformation (Eq. A3) is consistent with the transformation from Harris et al. (1991) and the zeropoint from Faber (1992) to $\sim 0.03$ magnitudes. This difference is well within the uncertainty of $\sim 0.1$ magnitudes for the $F702W$ zeropoint (Faber 1992).

The 9″ diameter aperture magnitudes have the advantage that no aperture correction is needed to compare them to the standard star measurements, making them suitable for the derivation of Eq. A3. However, they have large errorbars.

In the application of eq. A2 to our galaxy photometry, we used $R - I$ colors measured through 5″ apertures. These colors have much smaller errors. The measured values, corrected for galactic extinction, are listed in Table 2.

### A2 Correction to Redshifted $V$ Magnitude

The resulting $R$ and $I$ magnitudes can be related to the magnitude through a redshifted $V$ filter in the following way. Each magnitude is interpreted as a fluxdensity at the central frequency $\nu_i$ : $m_i = c_i - 2.5 \log F(\nu_i)$ The conversion constants are taken from Frei & Gunn (1994). Their constants are based on broad band observations of the spectrophotometric standards of Oke & Gunn (1983). The uncertainty in the constants is determined by the uncertainty in the spectrophotometry.

Next we assume that the flux density at the redshifted $V$ band is related to the flux densities in the $R$ and $I$ band by $F(\nu_V(z)) = F(\nu_R)^\alpha F(\nu_I)^{1-\alpha}$. It is straightforward to derive the magnitude in the redshifted $V$ band system

$$V_z = I + \alpha(R - I) + c_I + \alpha(c_R - c_I) - c_V + 2.5\log(1 + z) (A5)$$

The last term $2.5 \log(1 + z)$ makes the magnitude behave as if it is a flux, and not a fluxdensity. This implies that the surface brightness decreases as $(1 + z)^{-4}$, and not as $(1 + z)^{-3}$ as for surface brightness flux densities. The corrected magnitude $V_z$ behaves therefore as $K$-corrected magnitude.

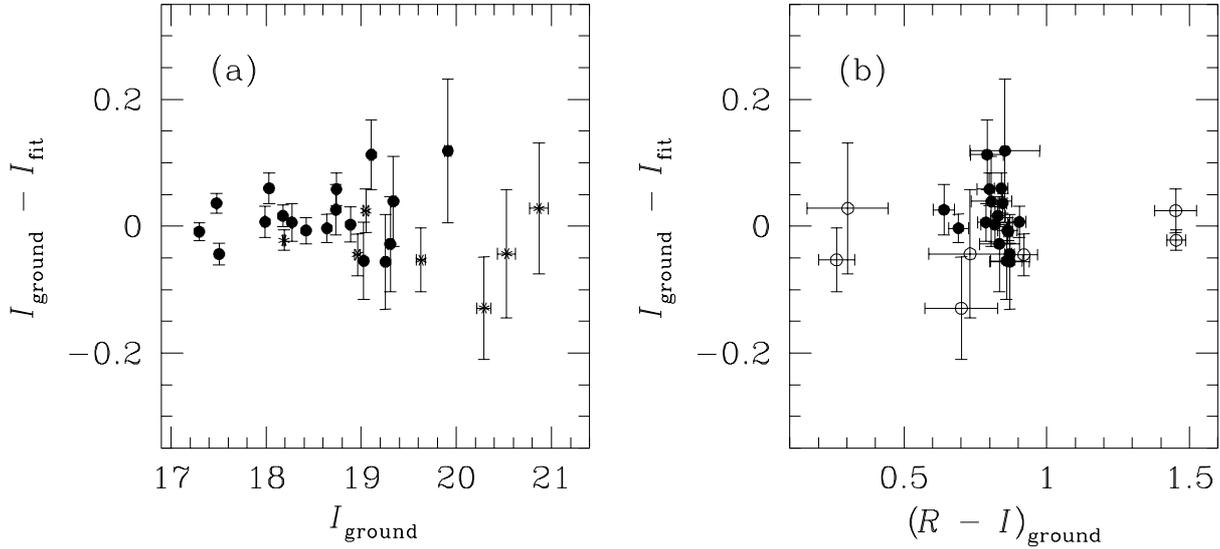

**Figure A1.** Comparison of ground based $I$ magnitudes with $I$ band magnitudes calculated from HST data using $I_{\rm fit} = F702W - 0.70(R - I - 0.82) - 2.92$. Filled circles indicate galaxies, open circles indicate stars. In (a), the difference $I_{\rm ground} - I_{\rm fit}$ is plotted against ground based $I$. In (b), $I_{\rm ground} - I_{\rm fit}$ is plotted against $R - I$ color. No systematic trends are apparent. The scatter is consistent with the measurement errors.

We calculated $\alpha$ for galaxy spectral energy distributions from Coleman et al. (1980) and Pence (1976), for $z = 0.39$. The value of $\alpha$ is $0.23 \pm 0.01$ for early-type galaxies and early-type spirals. The resulting transformation

$$V_z = I + 0.23(R - I) + 0.69 \qquad (A6)$$

if correct for all galaxy types to 0.01 magnitude. The most significant error in the transformation comes from the uncertainty in the spectrophotometry. We estimate this uncertainty at 0.03 from a comparison of synthetic $V - I$ colors of stars observed by both Oke (1990) and Hamuy et al. (1994).

## APPENDIX B: THE COMA DATA

The low redshift comparison sample consists of E and S0 galaxies in the Coma cluster at $z = 0.023$. The surface photometry is from Jørgensen, Franx & Kjærgaard (1995a) [JFK95a]. The data were taken in the Johnson $B$ and the Gunn $r$ photometric filters. The effective radii and surface brightnesses in $V$ were calculated from the $B$ band photometry, and $B - V$ colors. The $B$ band is the best band to use for this purpose, as the $V$ band is closer to the $B$ band than the Gunn $r$ band. Color gradients in $B - V$ are very small, and can be ignored.

The $B - V$ color was calculated from the $B - r$ color. The transformation is given by Jørgensen (1994):

$$B - V = 0.673(B - r) + 0.184 \qquad (\sigma_{\rm fit} = 0.021). \qquad (A7)$$

The $B - r$ colors were taken from Jørgensen et al. (1992), and seeing corrected as described in JFK95a.

JFK95a listed the mean surface brightness within an effective radius. These are $K$-corrected, and corrected for cosmological surface brightness dimming. We removed the correction for surface brightness dimming, and converted the surface brightnesses to the surface brightnesses at an $r_e$. The typical error in the $V$ surface brightness is 0.03 mag.

The velocity dispersions listed by JFK95b were used. These are based on velocity dispersions published by Davies et al. (1987), Dressler (1987), Lucey et al. (1991), and Guzmán et al. (1992). JFK95b corrected these dispersions to an aperture size of $3''\!.4$.